\documentclass{article}
\pdfoutput=1
\usepackage{arxiv}

\usepackage[utf8]{inputenc} 
\usepackage[T1]{fontenc}    
\usepackage{hyperref}       
\usepackage{url}            
\usepackage{booktabs}       
\usepackage{amsfonts}       
\usepackage{nicefrac}       
\usepackage{microtype}      
\usepackage{lipsum}

\usepackage{amsmath}
\usepackage{graphicx,psfrag,epsf}
\usepackage{enumerate}
\usepackage{natbib}

\usepackage{slashbox}

\newcommand{\Bbeta}{\boldsymbol{\beta}}

\newcommand{\Beta}{\boldsymbol{\beta}}

\newcommand{\BZ}{\boldsymbol{Z}}

\newcommand{\Bzeta}{\boldsymbol{\zeta}}
\newcommand{\BA}{\boldsymbol{A}}
\newcommand{\BB}{\boldsymbol{B}} 
\newcommand{\Bmu}{\boldsymbol{\mu}}
\newcommand{\BQ}{\boldsymbol{Q}}
\newcommand{\BGamma}{\boldsymbol{\Gamma}}

\title{\bf Quantile regression on inactivity time}

\author{
Lauren C. Balmert\thanks{Corresponding Author: lauren.balmert@northwestern.edu}\\
Department of Preventive Medicine (Biostatistics)\\
Feinberg School of Medicine\\
Northwestern University, Chicago, USA\\
\And
Ruosha Li\\
Department of Biostatistics and Data Science\\
University of Texas Health Science Center, Huston, USA\\
\And
Limin Peng\\
Department of Biostatistics and Bioinformatics\\
Rollins School of Public Health\\
Emory University, Atlanta, USA\\
\And
Jong-Hyeon Jeong \\
Department of Biostatistics\\
Graduate School of Public Health\\
University of Pittsburgh, Pittsburgh, USA\\
}

\begin{document}
	\maketitle

\begin{abstract}
The inactivity time, or lost lifespan specifically for mortality data, concerns time from occurrence of an event of interest to the current time point and has recently emerged as a new summary measure for cumulative information inherent in time-to-event data. This summary measure provides several benefits over the traditional methods, including more straightforward interpretation yet less sensitivity to heavy censoring. However, there exists no systematic modeling approach to inferring the quantile inactivity time in the literature. In this paper, we propose a regression method for the quantiles of the inactivity time distribution under right censoring. The consistency and asymptotic normality of the regression parameters are established. To avoid estimation of the probability density function of the inactivity time distribution under censoring, we propose a computationally efficient method for estimating the variance-covariance matrix of the regression coefficient estimates. Simulation results are presented to validate the finite sample properties of the proposed estimators and test statistics. The proposed method is illustrated with a real dataset from a clinical trial on breast cancer.
\end{abstract}

\section{Introduction} 

Time-to-event data can be encountered in many research areas such as engineering, economics, medicine, and social sciences. Statistical methods to analyze time-to-event data mainly utilize cumulative information up to the time of analysis while there also exists a long history of statistical methods for residual information such as mean residual life (Cs\"org\"o and Cs\"org\"o, 1987), or life expectancy (Deevey, 1947). On the other hand, inactivity time (Nanda et al., 2003; Li and Lu, 2003), also known as reversed residual life, has recently emerged as a new summary measure for cumulative information inherent in censored time-to-event data under the name of lost lifespan or life lost specifically for mortality data (Balmert and Jeong, 2016). Earlier Andersen (2013) defined ``years lost'' as a subtraction of the restricted {\it mean} lifetime from a prespecified time point under competing risks and extended it to a regression setting. The concept of inactivity time can be broadly applied to many research areas that involve time-to-event data such as survival analysis, reliability, and engineering. However, no systematic modeling approach exists to infer the quantile inactivity time under censoring, adjusting for confounding factors, in the literature. 

When the primary outcome for a study is time to an event of interest, the popular hazard rates or survival probabilities may be compared between groups based on the cumulative information, say, {\it up to year 10}. On the other hand, the distributions of the remaining years {\it beyond year 10} may be also compared, which might be an alternative summary measure with more intuitive and straightforward interpretation, but they can be heavily influenced by censored observations toward the tail of the distribution. The inactivity time is defined as the time lost due to an event that has occurred {\it before} a given time point (see Figure 1). A clear distinction between residual lifetime and inactivity time is that the subgroup of subjects targeted by the inactivity time analysis consists of subjects who do not survive up to time $t_0$, while the subgroup addressed by the residual lifetime only consists of subjects who survive beyond time $t_0$. Therefore, major advantages of using the inactivity time, or life lost,  in survival analysis would be that (i) it is a new way of summarizing time-to-event data in terms of lifetime lost rather than using the hazard function, survival probability and its inverse as quantiles, or residual life and (ii) provides straightforward interpretation as it has a time dimension like days, weeks, or years lost before a given time point rather than more mathematical quantities like the hazard function defined as the limiting conditional probability of instantaneous failure rate or other limiting concepts of probability. 

The quantity of life lost may be carefully interpreted in two ways; (i) for data analysis and (ii) for prediction. First, for the purpose of data analysis, suppose a clinical trial on a disease was performed, and data were collected on various patient characteristics, together with treatment group, time to an event of interest, and event status for a study period. To analyze this type of observed data set with some data points being right censored, the proposed quantile regression can provide a panoramic view of the treatment effect on years lost due to the event of interest, adjusted for some confounding factors, as the conditioning time $t_0$ progresses. In case of prediction, however, more care is needed since we are conditioning on a future time point of $t_0$. For example, it can be stated carefully such as ``If a patient fails within $t_0$ years after diagnosis of a disease, the median of the distribution of years lost would be $s$ years, so that if the patient gets treated with this medicine, it would decrease the years lost by $r$ years". Of course, the value of $t_0$ can vary for different practical scenarios, or a careful residual life analysis can be used in parallel to estimate a minimum value of $t_0$ as the estimated median residual lifetime of that particular patient.

Another potential application of the proposed model might be to epigenetic data from studies on biological age where subjects are treated with medications and followed until the end of the study period, at which time subjects' genomes are measured to evaluate how many days, weeks, and years were reversed in biological clock (Fahy {\it et al.}, 2019). This type of data would be different from the usual survival data in that events are not occurring as time progresses, but identified at the end of follow-up period as {\it reversed biological clock}. To maintain the feature of survival data, there could still be censored observations due to lost to follow-up, in which case the only available information would be that the length of reversed biological age would not have reached the observed censoring time point from the last follow-up. Under the setting of the proposed model, time-to-event can be defined as time from study entry to the time point where the age reverse has reached, and inactivity time can be renamed as reversed lifetime in this case.

\begin{figure}
	\includegraphics[width=0.6\linewidth]{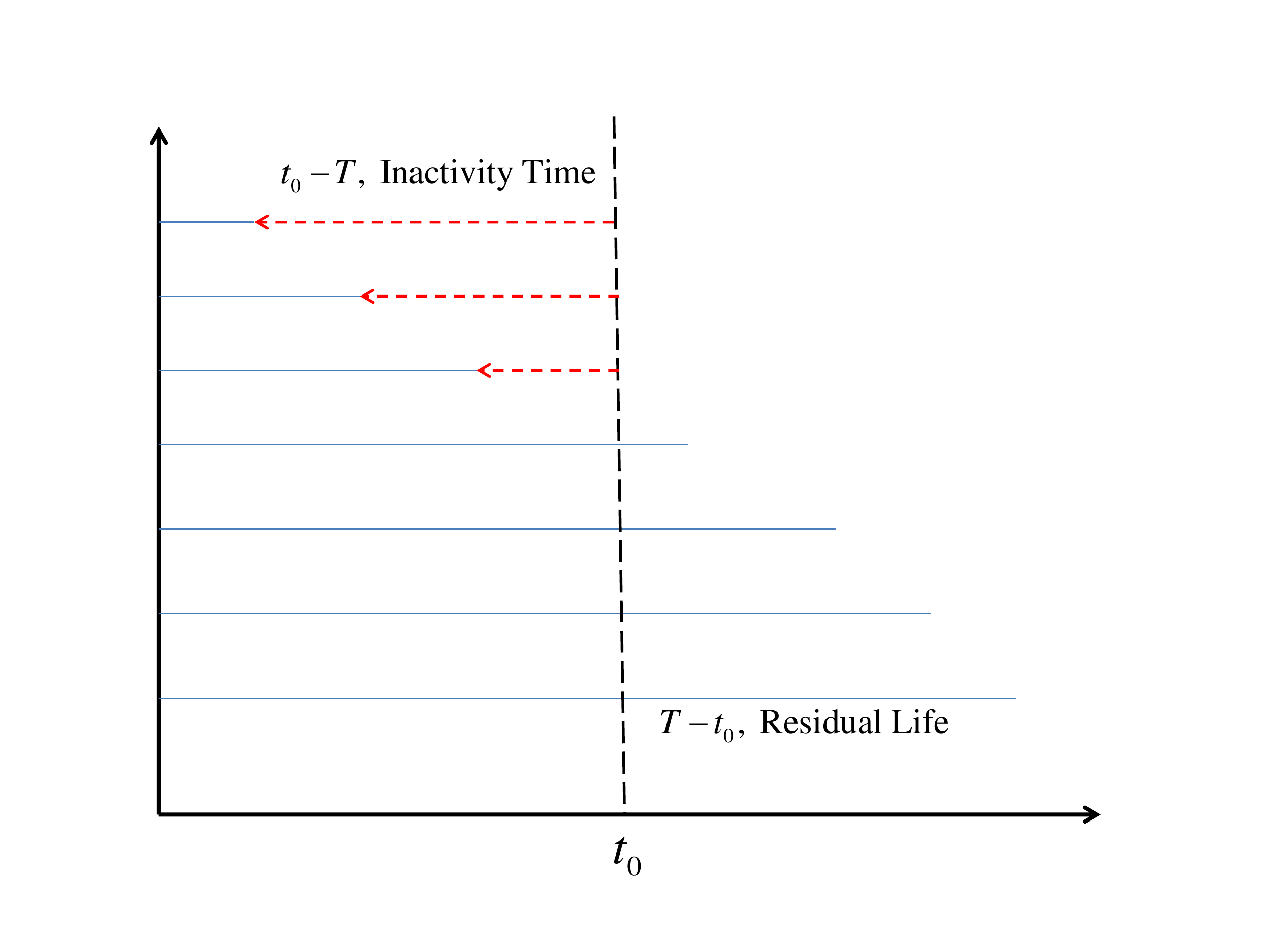}
	\centering
	\caption{Description of inactivity time (dashed lines with left arrows) and residual lifetime at a fixed time point $t_0$}
	\label{fig:figure1}
\end{figure}

Quantile regression, originally developed by Koenker and Basset (1978), is a well-studied extension of the linear regression (Portnoy and Koenker, 1997). Methods have been also  established for time-to-event data in the presence of censoring (Ying et al., 1995; Lindgren, 1997; McKeague et al., 2001; Yin and Cai, 2005; Peng and Huang, 2008). More recently, covariate effects on residual life were examined under the parametric proportional hazards and accelerated life models (Rao et al., 1992), and Bayesian modeling was also considered on the median residual life (Gelfand and Kottas, 2003). Jung et al. (2009) developed a method of the quantile regression on residual life, which has been extended to cause-specific quantile residual life regression (Lim and Jeong, 2015), and more recently to methods allowing for dynamic predictions (Li et al., 2016). In this paper, we propose a regression method on the quantiles of the distributions of the inactivity time adjusting for potential confounding factors under right censoring. 

In Section 2, we define the quantile inactivity time function and provide the notation to be used throughout the paper. In Section 3, proposed are an estimating equation, variance estimator, and test statistics for the regression parameters. In Section 4, the proposed method is assessed via simulation studies, which is applied to a breast cancer dataset in Section 5. Finally, we provide concluding remarks in Section 6.  

\section{Quantile Inactivity Time Function}
Throughout the paper, $T_{i}$ and $C_i$ will denote the potential event time and censoring time for the $i^{th}$ subject with survival functions of $S(t) = P(T_{i} \geq t)$ and  $G(t) = P(C_i\geq t)$, respectively. The random variable $Y_{i}$ will represent the observed survival time as the minimum of $T_{i}$ and $C_{i}$, and $\Delta_{i}$ will be an event indicator $(\Delta_{i} = 1$ if $Y_{i} = T_{i})$. The censoring distribution to be used in our estimating equation later will be estimated by the Kaplan-Meier estimator (Kaplan and Meier, 1958) denoted $\hat{G}(t)$. We will assume independence between $T_i$ and $C_i$. 

The inactivity time, defined specifically as lost lifespan for mortality
 data in Balmert and Jeong (2016), considers the time lost due to an event occurring prior to a specified time point, $t_{0}$. Here $t_{0}$ can be chosen such that the cumulative information up to $t_{0}$ can be statistically meaningful in terms of number of events as well as clinically meaningful in terms of milestones during the disease treatment period such as 5-, 10-, or, 15-year cancer-free survival as in the routine time-to-event analysis. 

Let us define the $\lambda$-percentile of the inactivity time distribution as
$$
\theta_{\lambda|t_0} = \lambda\mbox{-percentile} (t_{0} - T_{i} | T_{i} \leq t_{0} ).
$$

Then $\theta_{\lambda|t_0}$ satisfies $P(t_{0}-T_{i} \leq \theta_{\lambda|t_{0}} | T_{i} \leq t_{0}) = \lambda,$ or equivalently

 $$\frac{P(T_{i} \geq t_{0} -\theta_{\lambda|t_{0}}) - P(T_{i} > t_{0})}{1-P(T_{i} > t_{0})} = \lambda,$$
which can be rewritten in terms of the survival function as
$$ \frac{S(t_{0} - \theta_{\lambda|t_{0}}) - S(t_{0})}{1 - S(t_{0})} = \lambda. $$ 
Here given observed data and $\lambda$, $\theta_{\lambda|t_{0}}$ can be nonparametrically estimated after replacing ${S}(t)$ with its consistent estimator $\hat{S}(t)$ (Balmert and Jeong, 2017). Practically, in clinical intervention studies, researchers might be interested in knowing what would be a robust measure of the center of the distribution of life years the patients lost due to their deaths given the data up to 5 years. These measures can be also compared among intervention groups to infer the intervention effect of a study drug with or without adjusting for confounding factors. The purpose of this paper is to propose the following log-linear quantile regression model for inactivity time to $t_0$:
\begin{equation}
\lambda \mbox{-percentile}\{\ln(t_{0}-T_{i}) | T_{i} \leq t_{0}, \BZ_{i}\} = \Bbeta'_{\lambda|t_{0}}\BZ_{i}, \label{eqn;3.1}
\end{equation}
where $\Bbeta'_{\lambda|t_{0}}$ is a vector of the regression coefficients, $(\beta_{\lambda|t_{0},0},\beta_{\lambda|t_{0},1},...,\beta_{\lambda|t_{0},p})'$, and $\BZ_{i}$ is a vector of covariates for the $i^{th}$ individual, $(1, Z_{1i},...,Z_{pi})$. Here the regression parameter can be interpreted as the difference of the two quantile inactivity times on a log-scale when the corresponding covariate is binary. For a continuous covariate, it can be interpreted as an increment or decrement of the quantile inactivity time on a log-scale when the associated covariate increases by one unit. 

\section{Estimation and Inference}

Since model (\ref{eqn;3.1}) implies 
$$\lambda = P(t_{0}-T_{i} \leq \exp(\Bbeta_{\lambda|t_{0}}'\BZ_{i})|T_{i}\leq t_{0}),$$
we have
$$E[I(t_{0}  - T_{i} \leq \exp(\Bbeta_{\lambda|t_{0}}'\BZ_{i}) , T_{i} \leq t_{0} ) - \lambda I(T_{i} \leq t_{0})]=0.$$ 
Assuming conditional independence between $T_i$ and $C_i$ given $\BZ_{i}$ and the independence between $C_i$ and $\BZ_{i}$, which often occurs under administrative censoring in randomized clinical trials, it holds that
\begin{eqnarray}
&&E\left[\frac{I(T_{i} \geq t_{0} - \exp(\Bbeta_{\lambda|t_{0}}'\BZ_{i}), T_{i} \leq t_{0}, T_{i} \leq C_i)}{G(T_{i})}\right] \nonumber \\
&=& E\left[I\{T_{i} \geq t_{0} - \exp(\Bbeta_{\lambda|t_{0}}'\BZ_{i}) , T_{i} \leq t_{0}\}\right].\label{eqn;3.21}
\end{eqnarray}
Note that the independence assumption between $C_i$ and $\BZ_{i}$ can be relaxed so that $G(T_{i})$ can be replaced by $G(\cdot)$ through some additional regression modeling of $T_i$ given $Z_i$. Similarly we have
\begin{equation}
E\left[ \frac{I(T_{i} \leq t_{0} , T_{i} \leq C_i)}{G(T_{i})} \right] = E\left[I(T_{i} \leq t_{0})\right]. \label{eqn;3.22}
\end{equation}

Now that, given observed data, the events $\{T_i \geq t_0-\exp(\Bbeta_{\lambda|t_{0}}'\BZ_{i}),T_i \leq t_0, T_i \leq C_i\}$ and $\{T_{i} \leq t_{0} , T_{i} \leq C_i\}$ are equivalent to $\{t_0-\exp(\Bbeta_{\lambda|t_{0}}'\BZ_{i}) \leq Y_i \leq t_0, \Delta_i=1\}$ and $\{Y_i \leq t_0, \Delta_i=1\}$ respectively, equations (\ref{eqn;3.21}) and (\ref{eqn;3.22}) imply
$$E\left[\frac{\BZ_{i}I(Y_{i} \leq t_{0}, \Delta_{i} = 1)}{G(Y_{i})} \times [I\{t_{0} - Y_{i} \leq \exp(\Bbeta_{\lambda|t_{0}}'\BZ_{i})\} - \lambda ] \right] = 0. $$ 
Therefore the regression parameter $\Bbeta_{\lambda|t_{0}}$ can be estimated from the following equation under right censoring:
\begin{equation}
\BQ_n(\Bbeta_{\lambda|t_{0}}) = n^{-1/2} \sum_{i=1}^{n} \BZ_{i} \left[\frac{I(Y_{i} \leq t_{0}, \Delta_{i} = 1)}{\hat{G}(Y_{i})} \right] \times \left[\lambda-I\{\log(t_{0} - Y_{i}) \leq \Bbeta_{\lambda|t_{0}}'\BZ_{i}\} \right] \approx0, \label{eqn;3.2}
\end{equation}
where $\hat{G}(Y_{i})$ is the Kaplan-Meier estimate (Kaplan-Meier, 1958) of the censoring distribution based on the observed data ($Y_i,1-\Delta_i)$, assuming $C_i$ and $\BZ_i$ are independent. Note that the equation (\ref{eqn;3.2}) is the estimating equation for the weighted quantile regression with the weight function of $w_i=I(Y_{i} \leq t_{0}, \Delta_{i} = 1)/\hat{G}(Y_{i})$. More specifically, an individual term in the estimating equation (\ref{eqn;3.2}) takes the form of $w_i \BZ_{i} \psi_{\lambda}(u_i)$, where $\psi_{\lambda}(u_i)=\lambda-I(u_i < 0)$ and $u_i=\log(t_{0} - Y_{i})-\Bbeta_{\lambda|t_{0}}'\BZ_{i}$, which is the first derivative of the check function, i.e. $\rho_{\lambda}(u_i)=u_i(\lambda-I(u_i<0)$. The check function can be minimized by using a linear programming simplex-based method such as Barrodale-Roberts algorithm (Barrodale and Roberts, 1973), which was implemented as the default in the function \verb|rq()| with the \verb|weight| option in the R package \verb|quantreg|.

Suppose $\Bbeta^{0}_{\lambda|t_{0}}$ is the true value in the interior of a bounded convex region. Define 
$$\Bzeta_{1i}(\lambda)\equiv\BZ_i{I(Y_i\leq t_0,\Delta_i=1)}/{{G}(Y_i)}\left[I\big\{\log(t_0-Y_i)\leq \BZ_i'\Bbeta^0_{\lambda|t_0}\big\}-\lambda\right],$$
and
$$\Bzeta_{2i}(\lambda)=E_{\mathcal{D}_j}\left[-IF_i(Y_j)\BZ_j{I(Y_j\leq t_0,\Delta_j=1)}G^{-2}(Y_j)\left[I\big\{\log(t_0-Y_j)\leq \BZ_j'\Bbeta^0_{\lambda|t_0}\big\}-\lambda\right]\bigg|\mathcal{D}_i\right],$$
where $\mathcal{D}_i=(Y_i,\Delta_i,\BZ_i)$ denotes all observed data from the $i^{th}$ subject and 
$$IF_i(t)=G(t)\int_0^tr(s)^{-1}dM_i^G(s),$$ 
where $r(t)$ is the limiting value of the risk process for all subjects and $M_i^G(t)$ is the martingale process of the censoring time for the $i^{th}$ subject. Also define $\BB(\Bbeta^0_{\lambda|t_0})=\partial \Bmu(\Bbeta_{\lambda|t_0})/\partial \Bbeta_{\lambda|t_0}|_{\Bbeta_{\lambda|t_0}=\Bbeta^0_{\lambda|t_0}},$ where
$$\Bmu(\Bbeta_{\lambda|t_0})=E\left[\frac{\BZ N(t_0)}{G(Y)} [I\{Y \geq t_0-\exp(\BZ' \Bbeta_{\lambda|t_0})\}-\lambda] \right],$$
and $N(t)=I(Y \leq t, \Delta=1)$.
Then the following theorem states the uniform consistency of $\hat{\Bbeta}_{\lambda|t_{0}}$ and asymptotic normalities of the estimating equation (\ref{eqn;3.2}) and $\hat{\Bbeta}_{\lambda|t_{0}}$.

\vspace{.3cm}

\begin{theorem} Under the regularity conditions that (i) there exists $\tau>0$ such that $P(C_i=\tau)>0$ and 
$P(C_i>\tau)=0$, (ii) $P\{\log(t_{0}-T_{i}) \leq \tau\} > 0$ (Peng and Fine, 2009) and $\BZ' \Bbeta_{\lambda|t_{0}} \leq \tau$ with probability 1, and (ii) the conditional probability density function of $T_i$ given $\BZ_i$ is uniformly bounded,  
\begin{enumerate}
	\item $\hat{\Bbeta}_{\lambda|t_{0}} \rightarrow \Bbeta^{0}_{\lambda|t_{0}}$, a.s. as $n \rightarrow \infty$.
	\item $\BQ_n(\Bbeta^0_{\lambda|t_0})$ follows a zero-mean Gaussian process in $\lambda$, with the variance-covariance matrix of $\BGamma_{\lambda|t_{0}}=E[\Bzeta_i(\lambda)\Bzeta_i^{'}(\lambda)],$ where $\Bzeta_i(\lambda)=\Bzeta_{1i}(\lambda)+\Bzeta_{2i}(\lambda).$
	\item $\sqrt{n}\left(\hat{\Bbeta}_{\lambda|t_0}-\Bbeta^0_{\lambda|t_0}\right)$ weakly converges to a Gaussian process with the variance-covariance matrix of 
\begin{equation}
\BB(\Bbeta^0_{\lambda|t_0})^{-1}E[\Bzeta_i(\lambda)\Bzeta_i^{'}(\lambda^{*})] \BB(\Bbeta^0_{\lambda|t_0})^{-1'}, \label{eqn;3.3}\end{equation}
\end{enumerate}
where $\lambda^{*}$ is a different value of $\lambda$.
\end{theorem}

Assumptions in (i) are often satisfied in the presence of adminstrative censoring. In the general case, $C$ can be truncated by $\min(C, L)$, with $L$ being chosen as a constant slightly smaller than the observed upper bound of $C$'s support, in which case those assumptions hold. As long as $L$ is only slightly smaller than $C$, we expect truncating $C$ would incur very minimal information loss.

The form of the variance-covariance matrix in (\ref{eqn;3.3}) is different from those of the standard likelihood theory or the Cox's partial likelihood approach, where the inverse of the negative Hessian matrix is the corresponding variance-covariance matrix.

Under the regularity conditions (i)-(iii), we prove the consistency of $\hat{\Bbeta}_{\lambda|t_{0}}$ in Appendix A, and establish the asymptotic normalities of the estimating equation $\BQ_n(\Bbeta^0_{\lambda|t_0})$ and the proposed estimator $\hat{\Bbeta}_{\lambda|t_{0}}$ in Appendix B.

Under the null hypothesis of $H_{0}: \Bbeta_{\lambda|t_{0}} = \Bbeta_{\lambda|t_{0},0}$, a test statistic for the global test can be constructed based on the asymptotic distribution of the estimating function $n^{-1/2} \BQ_n(\Bbeta_{\lambda|t_{0}})$ in (\ref{eqn;3.2}) as 
	 $$n^{-1}\BQ^{'}_{n}(\Bbeta_{\lambda|t_{0},0})\hat{\BGamma}^{-1}_{\lambda|t_{0}}\BQ_n(\Bbeta_{\lambda|t_{0},0}),$$
which approximately follows a $\chi^{2}$-distribution with $p + 1$ degrees of freedom, where $p$ is the number of covariates. However, a test statistic for a subset of $\Bbeta_{\lambda|t_{0}}$, e.g. $\Bbeta^{(1)}_{\lambda|t_{0}}$, would also include the remaining parameters not being tested as nuisance parameters. A variation of the minimum dispersion statistic (Basawa and Koul, 1988) can be adopted to eliminate the nuisance parameters, but the computational burden could be enormously heavy especially when a large number of regression coefficients are included in the model.
	
For this reason and also to avoid estimation of the probability density function of $(t_0-T_i)I(T_i < t_0)|Z_i$ under censoring, we have employed a perturbation method (Jin {\it et al.}, 2001) to estimate the limiting distribution of $\hat{\Bbeta}_{\lambda|t_{0}}$, from which confidence intervals could be obtained using the normal approximation of $\hat{\Bbeta}_{\lambda|t_{0}}$. Specifically, the weight function in the estimating equation (\ref{eqn;3.2}) was perturbed by a set of independent random variates from the unit exponential distribution, i.e. $(\xi_1,\xi_2,...,\xi_n)$, and the regression parameters $\Bbeta^*_{\lambda|t_{0}}$ were estimated from 
$$n^{-1/2}\sum_{i=1}^{n} \BZ_{i} \xi_i \left[\frac{I(Y_{i} \leq t_{0}, 
\Delta_{i} = 1)}{G^*(Y_{i})} \right] \times \left[I(\log(t_{0} - Y_{i}) \leq \Bbeta_{\lambda|t_{0}}^{* \prime} \BZ_{i}) - \lambda \right] \approx 0,$$
where $G^*(Y_i)$ is obtained from perturbing the indicator functions for the risk sets and the event indicators in $\hat{G}(Y_{i})$ by the same exponential variates. Given data, the random variates $(\xi_1,\xi_2,...,\xi_n)$ were repetatively generated and a large number of realizations of $\Bbeta^*_{\lambda|t_{0}}$ were obtained. Following the arguments of Jin et al. (2001), we can show that the conditional distribution of $n^{1/2}(\Bbeta^*_{\lambda|t_{0}}-\hat{\Bbeta}_{\lambda|t_{0}})$ given the observed data is asymptotically equivalent to the unconditional distribution of $n^{1/2}(\hat{\Bbeta}_{\lambda|t_{0}}-\Bbeta_{\lambda|t_{0},0})$ as a process of $\lambda$. For fixed $\lambda^*$, the variance-covariance matrix of $\hat{\Bbeta}_{\lambda|t_{0}}$ can be estimated by the sample variance-covariance matrix of $\Bbeta^*_{\lambda|t_{0}}$'s, which can be used to infer an individual or a subset of the regression coefficients.

\section{Simulation Studies}
Several simulation studies were performed to assess the performance of the proposed estimators and test statistics with finite samples. We generated data from a parametric proportional hazards model (Cox, 1972) with a Weibull distribution as the baseline distribution and one group indicator as a covariate. Thus, the true survival function follows 
\begin{equation}
S(t) = \exp(-(\rho t)^{\eta}\exp(\beta z_{i})), \label{eqn;4.1}
\end{equation}
where the Weibull parameters $\rho$ and $\eta$ are set to be 0.2 and 2, respectively, througout the simulation studies and $\beta$ is the regression parameter associated with the group indicator $z_{i}$ ($z_{i}=0$ for the control and $z_{i}=1$ for an intervention). Under the parametric Cox model (\ref{eqn;4.1}), the true median inactivity time equals 
\begin{equation}
\theta_{t_{0}}(z) = t_{0}-\frac{1}{\rho}[\exp(-\beta z)\{\log(2)-\log(1+\exp(-(\rho t_{0})^{\eta}\exp(\beta z))\}]^{1/\eta}. \label{eqn;4.2}
\end{equation}

Potential censoring times $C_i$ were generated from a uniform distribution on $[a,b]$, where $a$ and $b$ were chosen to render the desired censoring proportions. Observed survival times $Y_{i}$ were then determined as the minimum of potential failure times and potential censoring times, i.e. $\min(T_i,C_i)$.  

First, we evaluate the estimation performance of our proposed method. The true values of $\theta_{t_{0}}$ in (\ref{eqn;4.2}) when $\beta = 0$ would be the same for both control and intervention groups as 10.8, 9.8, 8.8, and 7.8 at $t_{0}$ = 15, 14, 13, and 12, respectively.  Let us consider a simple log-linear {\it median} regression model for inactivity time,
\begin{equation}
\mbox{med}(\ln(t_{0}-T_{i})|T_{i} \leq t_{0}) = \beta_{t_{0}}^{(0)} + \beta_{t_{0}}^{(1)}z_{1i}, \label{eqn;4.3}
\end{equation}
where $z_{1i}$ is a binary covariate indicating intervention group $(z_{1i}=1)$ or control group $(z_{1i}=0)$, and $\beta_{t_{0}}^{(0)}$ and  $\beta_{t_{0}}^{(1)}$ are the intercept and a regression coefficient associated with $z_{1i}$, respectively. Following the invariance property of the log-transformation, the model is equivalent to 
$$\mbox{med}(t_{0}-T_{i}|T_{i} \leq t_{0}) = \exp(\beta_{t_{0}}^{(0)} + \beta_{t_{0}}^{(1)}z_{1i}),$$
implying that $\exp(\beta_{t_{0}}^{(0)})$ and $\exp(\beta_{t_{0}}^{(0)}+\beta_{t_{0}}^{(1)})$ can be interpreted as the median inactivity time in the control group and in the intervention group, respectively. Thus, the difference in median inactivity times between two groups is given by $\exp(\beta_{t_{0}}^{(0)})(\exp(\beta_{t_{0}}^{(1)})-1)$, and the ratio of two inactivity times by $\exp(\beta_{t_{0}}^{(1)})$, so that testing a null hypothesis of $\beta_{t_{0}}^{(1)} = 0$ will be equivalent to testing whether the ratio of two median inactivity times equals 1. 

In order to evaluate our parameter estimates, we compare $\hat{\beta}_{t_{0}}^{(1)}$ to 0 and $\hat{\beta}_{t_{0}}^{(0)}$ to the logarithm of the true median inactivity time from (\ref{eqn;4.2}) under $H_0$. At time point 15, for example, the true median inactivity time of 10.8 corresponds to $\beta_{t_{0}}^{(0)}$ = 2.38 and $\beta_{t_{0}}^{(1)}$ = 0 under the simple log-linear regression model (\ref{eqn;4.3}). As described in Section 3, we estimated the regression coefficients using the \verb|rq()| function with the weight function of $I(Y_{i} \leq t_{0}, \Delta_{i} = 1) / \hat{G}(Y_{i})$. Then, we used the perturbation method to estimate the variance-covariance matrix of $\hat{\Bbeta}_{t_0}$ and construct confidence intervals for $\beta^{(1)}_{t_0}$ using the normal approximation. Four hundred (400) perturbations were implemented for each simulation. 

Table 1 displays the results based on 1000 simulations with 200 observations per group. The bias and standard deviation of the parameter estimates were used to evaluate the empirical distribution of $\beta_{t_{0}}^{(0)}$ and  $\beta_{t_{0}}^{(1)}$ given various $t_0$'s (15, 14, 13, and 12) and censoring proportions (10\%, 20\%, and 30\%). For each simulation, the SE's for the parameter estimates were calculalted from 400 perturbations, which were used to construct confidence intervals for the true parameters. The average of those 1,000 SE's are presented under the column of ``ASE". One can notice that the biases are minimal under all scenarios, and the ASE's are overall close to SD's. Table 1 also presents the median inactivity time estimates for control and intervention groups. As the censoring proportion increases, the differences between parameter estimates and their true values slightly increase. The empirical standard deviations also inflate as the censoring proportion increases and as $t_{0}$ decreases.    
 

\vspace{.5cm}

\begin{table}
	\centering
	\caption{Bias and standard deviation of the empirical estimates of true regression parameters $\beta_{t_{0}}^{(0)}$ = 2.38, 2.29, 2.18, and 2.06 and $\beta_{t_{0}}^{(1)}$=0 at $t_{0}$ = 15, 14, 13, and 12; $\hat{\theta}^{(0)}$, estimated median inactivity time in control group; $\hat{\theta}^{(1)}$, estimated median inactivity time in intervention group; c\%, censoring proportion }
	\begin{tabular}{c c c c c c c c c c } 
		\hline\hline 
		$t_{0}$ & $c\%$ & Bias($\hat{\beta}_{t_{0}}^{(0)}$) & SD($\hat{\beta}_{t_{0}}^{(0)}$) &ASE($\hat{\beta}_{t_{0}}^{(0)}$) & Bias($\hat{\beta}_{t_{0}}^{(1)}$) & SD($\hat{\beta}_{t_{0}}^{(1)}$) & ASE($\hat{\beta}_{t_{0}}^{(1)}$) & $\hat{\theta}^{(0)}$ & $\hat{\theta}^{(1)}$  \\
		\hline 
		15 & 10 & 0.0005 & 0.0291 &0.0296 & 0.0001 & 0.0428 &0.0438& 10.843 & 10.844   \\ 
		 & 20 & 0.0005 & 0.0314 &0.0312 & 0.0004 & 0.0451 &0.0466& 10.843 & 10.847  \\ 
		& 30 & -0.0012 & 0.0332 & 0.0340& 0.0011 & 0.0494 &0.0500& 10.825 & 10.837  \\   \hline 
		14 & 10 & 0.0015 & 0.0328 &0.0330 & -0.0019 & 0.0466 &0.0482& 9.853 & 9.834   \\  
		& 20 & 0.0007 & 0.0350 &0.0347 & 0.0006 & 0.0513 &0.0514& 9.845 & 9.851  \\ 
		& 30 & -0.0004 & 0.0365 & 0.0370& 0.0002 & 0.0561 &0.0552& 9.835 & 9.836  \\ \hline 
		13 & 10 & -0.0004 & 0.0351 &0.0361 & -0.0014 & 0.0519 &0.0530& 8.837 & 8.825   \\  
		& 20 & -0.0013 & 0.0359 &0.0389 & -0.0007 & 0.0532 &0.0577& 8.829 & 8.823   \\ 
		& 30 & -0.0021 & 0.0412 & 0.0414& 0.0004 & 0.0602 &0.0612& 8.822 & 8.826  \\ \hline 
		12 & 10 & -0.0004 & 0.0391 &0.0406 & -0.0010 & 0.0567 &0.0602& 7.843 & 7.836   \\ 
		& 20 & -0.0033 & 0.0427 &0.0431 & 0.0017 & 0.0625 &0.0646& 7.821 & 7.834   \\ 
		& 30 & -0.0003 & 0.0455 & 0.0467& -0.0005 & 0.0659 &0.0691& 7.844 & 7.840   \\
		\hline 
	\end{tabular}
\end{table}

We then assessed the proposed test statistic in terms of rejection probabilities of the null hypothesis of $H_{0}: \beta_{t_{0}}^{(1)}=0$ at a two-sided significance level of 0.05 for different values of $\beta_{t_{0}}^{(1)}$, given various $t_0$'s (15, 14, 13, and 12), censoring proportions (10\%, 20\%, and 30\%), and sample sizes (100 and 200). The rejection probability was calculated as the mean, over the 1,000 simulations, of the proportions that 95\% confidence intervals from 400 perturbations do not include the null value of $\beta_{t_{0}}^{(1)}=0$. Therefore, the column under $\beta_{t_{0}}^{(1)}=0$ in Table 2 displays type I error probability for testing the null hypothesis of $H_{0}: \beta_{t_{0}}^{(1)}=0$. For power analysis, we have generated data under the parametric proportional hazards model  in (\ref{eqn;4.1}) by increasing the value of $\beta_{t_{0}}$ to induce differences between control and intervention groups. We set the true coefficient $\beta_{t_{0}}=-0.44, -0.82$, and $-1.18$ in (\ref{eqn;4.2}), which is equivalent to increasing the differences in median inactivity time between control and intervention by 1, 2, and 3. The results are displayed in Table 2. Empirical type I error probabilities are generally close to 0.05 regardless of different censoring proportions or sample sizes. Power decreases as $t_{0}$ decreases since less observations are included in the analysis, and increases as $\beta_{t_{0}}$ decreases, indicating a greater power to detect a larger difference between groups. Power decreases slightly as the censoring proportion increases, but we still have reasonable power to detect small absolute differences under heavy censoring with a smaller sample size of 100. Power also increases as sample size increases, as expected. 

\begin{table}
	\centering
	\caption{Empirical rejection rates for values of $\beta_{t_{0}}^{(1)}$ }
	{\small
	\begin{tabular}{c| c| c c c c c c c c c c c c c c} 
		\hline 
		& & \multicolumn{4}{c}{$n=100$} & & \multicolumn{4}{c}{$n=200$} & & \multicolumn{4}{c}{$n=1000$}\\  \cline{2-6} \cline{8-11}  					\cline{13-16}
		$t_0$& \backslashbox{$c\%$}{$\beta_{t_{0}}^{(1)}$} & 0.0 & -0.44 & -0.82 & -1.18  & & 0.0 & -0.44 & -0.82 & -1.18  & & 0.0 & -0.44 & -0.82 & -1.18  \\
		\hline
		15 & 10 & 0.041 &0.363  & 0.823 & 0.969  & & 0.052 & 0.676 & 0.994 & 1.000 && 0.051 & 0.999 &1.000 & 1.000     \\ 
		& 20 & 0.036 &0.287 & 0.701 & 0.940 & & 0.043 &0.559 & 0.950 & 0.999   && 0.038 & 0.997 & 1.000 & 1.000  \\ 
		& 30 & 0.045 &0.210 & 0.614 & 0.854  & & 0.046 &0.417 & 0.842 & 0.999   && 0.055 & 0.981 & 1.000 & 1.000  \\  \hline 
		14 & 10 & 0.044 &0.351  & 0.804 & 0.946 & & 0.055 &0.670 & 0.992 & 1.000  && 0.054 & 0.999 & 1.000 & 1.000  \\ 
		& 20 & 0.040 &0.303 & 0.686 & 0.923  & & 0.036 & 0.589 & 0.936 & 1.000   && 0.049 & 0.999 & 1.000 & 1.000   \\ 
		& 30 & 0.038 &0.219 & 0.599 & 0.821  & & 0.039 &0.372 & 0.827 & 0.99   && 0.041 & 0.981 & 1.000 & 1.000  \\  \hline 
		13 & 10 & 0.048 &0.357 & 0.792 & 0.936  & & 0.046 &0.639 & 0.981 & 0.999    && 0.050 & 0.999 & 1.000 & 1.000 \\ 
		& 20 & 0.041 &0.262 & 0.662 & 0.885  & & 0.034 &0.580 & 0.926 & 0.998   && 0.041 & 0.997 & 1.000 & 1.000   \\ 
		& 30 & 0.043 &0.177 & 0.567 & 0.783  & & 0.041 &0.398 & 0.801 & 0.985   && 0.041 & 0.977 & 1.000 & 1.000  \\  \hline 
		12 & 10 & 0.032 &0.340 & 0.749 & 0.888 & & 0.044 &0.597 & 0.977 & 1.000  && 0.047 & 0.999 & 1.000 & 1.000    \\ 
		& 20 & 0.039 &0.294 & 0.597 & 0.830  & & 0.034 &0.536 & 0.886 & 0.992  && 0.043 & 0.977 & 1.000 & 1.000    \\ 
		& 30 & 0.044 &0.174 & 0.523 & 0.737 & & 0.039 &0.367 & 0.766 & 0.977   && 0.051 & 0.967 & 1.000 & 1.000  \\
		\hline 
	\end{tabular}
	}
\end{table}
\section{Application}

In this section, we apply the proposed estimation procedure and test-statistic to a real dataset from a clinical trial on breast cancer, i.e. NSABP (National Surgical Adjuvant Breast and Bowel Project) B-04 dataset (Fisher et al. 2002), which contains survival information on 1,665 breast cancer patients. The primary outcome of interest in this analysis is time to death. In addition to follow-up information, surgery type, and nodal status, the dataset also contains other covariates including age at diagnosis and pathological tumor size. In our analysis, we consider the following covariates: nodal status as a binary covariate with 0 for node-negative and 1 for node-postive, and both age at diagnosis and pathological tumor size as continuous covariates. There were 1,079 node-negative women and 586 node-positive women. Age at diagnosis ranged from 20 to 87 years with the mean of 55.4, and pathological tumor size ranged from 0 to 250mm with the mean of 34.1mm. Additionally, the median follow-up was 26 years with the overall censoring proportion of 23\%. In the models, the continuous covariates were multiplied by 0.01, for computational convenience. In our analysis, the main interest is how many more years the node-positive patients are expected to lose compared to the node-negative patients at various time points after surgery, adjusted for age at diagnosis and tumor size. In this particular cancer mortality dataset, the inactivity time, specifically referred to as lost lifespan in this section, is defined as the number of years lost due to death following a surgery. Our goal is to infer the effects of covariates on the median (or a quantile) of the lost lifespan distribution, and predict the median lost lifespan adjusting for significant covariate effects.

First, we used the proposed method to evaluate the significance of nodal status in the univariate log-linear quartile regression model (\ref{eqn;3.1}) ($\lambda=0.25, 0.5, 0.75$) that only includes nodal status as a covariate.  The test statistic was calculated at 3 time points ($t_{0}$ = 15, 20, and 25 years after surgery). Table 3 summarizes the results, including the parameter estimates $\hat \beta^{(intercept)}$ for the intercept and $\hat \beta^{(node)}$ for the effect of nodal status, and their 95\% confidence intervals calculated from the perturbation method. Significance of the nodal status parameter was indicated by a 95\% confidence interval not containing 0. Note that regardless of different time points specified, the quartile lost lifespans were significantly different between the two nodal groups. The node positive group had consistently longer quartile lost lifespans across all time points indicating worse prognosis in survival. The difference between nodal status groups also increased as time point increased or $\lambda$ decreased. The results from the simple log-linear median ($\lambda=0.5$) regression model presented here are also consistent with the ones from the two-sample test statistic proposed in Balmert and Jeong (2016).     

\vspace{.5cm}

\begin{table}
	\centering
	\caption{Parameter estimates and 95\% confidence intervals from the univariate log-linear quartile ($\lambda=0.25, 0.5, 0.75$) regression models}
	\begin{tabular}{c c c c c } 
		\hline 
		$\lambda$ & $\hat{\Bbeta}$ & $t_{0}$ = 15 & $t_{0}$ = 20 & $t_{0}$ = 25 \\
		\hline
		0.25 & $\hat{\beta}^{(intercept)}$ & 1.71 (1.60, 1.83) & 2.03 (1.90, 2.16) & 2.23 (2.07, 2.38)   \\ 	
		& $\hat{\beta}^{(node)}$ &  0.30 (0.15, 0.45) & 0.33 (0.16, 0.51) & 0.42 (0.22, 0.63)   \\ \hline	
		0.50 & $\hat{\beta}^{(intercept)}$ & 2.25 (2.20, 2.30) & 2.58 (2.54, 2.63) & 2.81 (2.76, 2.87)   \\ 	
		& $\hat{\beta}^{(node)}$ & 0.12 (0.04, 0.20) & 0.13 (0.06, 0.19) & 0.15 (0.08, 0.23)  \\ \hline  
		0.75 & $\hat{\beta}^{(intercept)}$ & 2.50 (2.47, 2.52) & 2.81 (2.79, 2.83) & 3.05 (3.03, 3.07)  \\
		& $\hat{\beta}^{(node)}$ & 0.07 (0.04, 0.11) & 0.07 (0.04, 0.10) & 0.07 (0.05, 0.10)  \\ 
	    \hline 
	\end{tabular}
\end{table}

Now we extend our analysis to a log-linear quartile regression model containing nodal status, age at diagnosis, and pathological tumor size as covariates. Using similar notations as before, let $\beta^{(age)}$ and $\beta^{(size)}$ denote the effects of additional covariates, age at diagnosis and pathological tumor size, respectively. Each covariate was tested separately for its significance using the confidence interval approach as previously described. The parameter estimates and corresponding 95\% confidence intervals are shown in Table 4. Except the median regression analysis at $t_0=15$, the nodal status remained statistically significant in all the other multivariate models. Additionally, the difference between node-negative and node-positive groups increased as $t_0$ increased, similarly to the results from the simple log-linear median regression models, except the 3$^{rd}$ quartile ($\lambda=0.75$) regression model. Age at diagnosis was mostly significant except the 1$^{st}$ quartile ($\lambda=0.25$) regression model at $t_0=25$ while pathological tumor size was consistently significant in all models. The proposed regression model allows for predicting a patient's median lost lifespan for a given time point based on significantly important factors, i.e. nodal status and age at diagnosis. For example, a 30-year old woman with positive lymph nodes and tumor size of 50mm is expected to have a median lost lifespan of 17.6 years ($=\exp\{2.86 + 0.08\times 1 -0.62 \times (0.01\times 30) + 0.24\times(0.01\times50)\})$ at 20 years after diagnosis. In comparison, a 30-year old patient with negative lymph nodes and tumor size of 50mm is expected to have a median lost lifespan of 16.3 years at 20 years after diagnosis.   

\begin{table}
	\centering
	\caption{Parameter estimates and corresponding 95\% confidence intervals from the multivariate log-linear quartile regression models ($\lambda=0.25,0.5,0.75$) using the proposed perturbation method}
	\begin{tabular}{c c c c c } 
		\hline 
		$\lambda$ & $\hat{\Bbeta}$ & $t_{0}$ = 15 & $t_{0}$ = 20 & $t_{0}$ = 25 \\
		\hline
		0.25 & $\hat{\beta}^{(intercept)}$ & 2.29 (1.87, 2.71) & 2.87 (2.15, 3.23) & 2.51 (1.95, 3.06)  \\ 	
		& $\hat{\beta}^{(node)}$ & 0.24 (0.09, 0.38) & 0.25 (0.09, 0.41) & 0.37 (0.15, 0.58)  \\  
		& $\hat{\beta}^{(age)}$ & -1.31 (-2.02, -0.60) & -1.55 (-2.12, -0.99) & -0.68 (-1.44, 0.08)  \\ 
		& $\hat{\beta}^{(size)}$ &  0.47 (0.19, 0.74) & 0.29 (-0.09, 0.67) & 0.36 (-0.04, 0.76)  \\ \hline	
		0.50 & $\hat{\beta}^{(intercept)}$ & 2.57 (2.40, 2.74) & 2.86 (2.69, 3.02) & 3.00 (2.87, 3.12)  \\ 
		& $\hat{\beta}^{(node)}$ & 0.06 (-0.01, 0.13) & 0.08 (0.01, 0.15) & 0.12 (0.07, 0.18)   \\ 	
		& $\hat{\beta}^{(age)}$ & -0.70 (-0.94, -0.46) & -0.62 (-0.85, -0.40) & -0.45 (-0.64, -0.25)  \\  	
		& $\hat{\beta}^{(size)}$ &	 0.24 (0.08, 0.39) & 0.24 (0.09, 0.39) & 0.24 (0.12, 0.36)  \\ \hline 
		0.75 & $\hat{\beta}^{(intercept)}$ &  2.55 (2.49, 2.62) & 2.87 (2.81, 2.93) & 3.11 (3.06, 3.17)  \\ 
		& $\hat{\beta}^{(node)}$ & 0.05 (0.02, 0.08) &  0.05 (0.02, 0.08) &  0.05 (0.02, 0.08)   \\  
		& $\hat{\beta}^{(age)}$ & -0.20 (-0.31, -0.09) & -0.18 (-0.27, -0.08) & -0.17 (-0.26, -0.08)   \\  
		& $\hat{\beta}^{(size)}$ & 0.16 (0.09, 0.22) & 0.13 (0.07, 0.19) & 0.12 (0.05, 0.19)  \\ 	
		\hline
		\hline
	\end{tabular}
\end{table}

\section{Conclusions}
The inactivity time, or lost lifespan specifically for mortality data, is a simple summary measure for time-to-event data that provides more straightforward interpretation yet is less sensitive to right  ensoring compared to residual life. In this paper, we proposed a new regression method for analyzing covariate effects on the quantiles of the distribution of inactivity time. Asymptotic properties were derived for the regression parameter estimators and test statistics. Simulation studies validated the estimation and inference procedure under various scenarios, and the proposed method was illustrated with an application to a breast cancer dataset. The proposed model does not have strong assumption like proportional hazards, and provides a new and sensible perspective to understand treatment effects or covariate effects, so that it can be a useful alternative in survival modeling.

Even if a direct comparison between the proposed model and the popular Cox's proportional hazards model would not be fair due to different model assumptions and simply because they are different summary measures of time-to-event data, both approaches could be useful for clinicians from different perspectives to communicate intervention options to patients. Another candidate model to be compared would be accelerated failure time (AFT) model (Kalbfleisch and Prentice, 2002), which is a log-linear model in failure time that is also different from our proposed model in this paper in both model assumptions and definitions of the summary measure. Possible extensions of the proposed model would be to include time-dependent covariates, competing risks, and random effects, which will merit future research. 

\section*{Acknowledgments}
Dr. Li's research was supported in part by NIH grant 1R01DK117209. Dr. Peng's research was supported in part by NIH grant R01HL-113548. Dr. Jeong's research was supported in part by National Institute of Health (NIH) grant 5-U10-CA69651-11.

\begin{center}
{\bf Appendix A: Consistency of $\hat{\Bbeta}_{\lambda|t_{0}}$}
\end{center}

We start by defining  
\begin{eqnarray*}
\tilde{\BQ}_{n}(\Bbeta_{\lambda|t_{0}}) 
&=& n^{-1/2}
\sum_{i=1}^{n} \BZ_{i}G(T_i)^{-1}[P\{T_{i} \geq t_{0}-\exp(\Bbeta'_{\lambda|t_{0}} \BZ_{i}),T_i \leq t_0, T_i \leq C_i\} \\
&& - \lambda P(T_i \leq t_0, T_i \leq C_i) ],
\end{eqnarray*}
which is equivalent to
\begin{eqnarray}
\tilde{\BQ}_{n}(\Bbeta_{\lambda|t_{0}}) 
&=&n^{-1/2}\sum_{i=1}^{n} \BZ_{i}G(Y_i)^{-1} [P\{t_{0}-\exp(\Bbeta'_{\lambda|t_{0}} \BZ_{i}) \leq Y_i \leq t_0,\Delta_i=1\} \nonumber \\
 && - \lambda P(Y_i \leq t_0,\Delta_i=1) ],\label{eqn:A0}
\end{eqnarray}
since the events $\{T_i \geq t_0-\exp(\Bbeta_{\lambda|t_{0}}'\BZ_{i}),T_i \leq t_0, T_i \leq C_i\}$ and $\{T_{i} \leq t_{0} , T_{i} \leq C_i\}$ are equivalent to $\{t_0-\exp(\Bbeta_{\lambda|t_{0}}'\BZ_{i}) \leq Y_i \leq t_0, \Delta_i=1\}$ and $\{Y_i \leq t_0, \Delta_i=1\}$, respectively, as introduced following the equation (\ref{eqn;3.22}) in Section 3.
When $\Bbeta_{\lambda|t_{0}}$ is replaced with $\Bbeta^{0}_{\lambda|t_{0}}$, the true value in the interior of a bounded convex region D, the above equation reduces to 0 approximately. Following Cs\"org\"o and Horv\'ath (1983), we know that for all $\epsilon > 0$, 
$$ \sup_{s\leq \tau} |\hat{G}(s)-G(s)| = o(n^{-1/2 + \epsilon}), \ a.s. $$
where $\tau$ is a constant satisfying $P\{\log(t_{0}-Y_{i}) \leq \tau\} > 0$ and $\Bbeta'_{\lambda|t_{0}}\BZ \leq \tau$, with probability 1.
This can be used to show that for $\Bbeta_{\lambda|t_{0}} \in D$, 
\begin{eqnarray*}
&&\sqrt{n}[\BQ_n(\Bbeta_{\lambda|t_{0}})-\tilde{\BQ}_{n}(\Bbeta_{\lambda|t_{0}})] \\
&=& \sum_{i=1}^{n} \BZ_{i} G(Y_i)^{-1} [I\{t_{0} - \exp(\Bbeta'_{t_{0}}\BZ_{i}) \leq Y_{i} \leq t_0, \Delta_{i}=1\} - P\{t_{0} - \exp(\Bbeta'_{t_{0}}\BZ_{i}) \leq Y_{i} \leq t_0, \Delta_{i}=1\}\\
&&-\lambda \{I(Y_i \leq t_0,\Delta_i=1)-P(Y_i \leq t_0,\Delta_i=1)\}]+o(n^{1/2 + \epsilon}), \ a.s.
\end{eqnarray*}
Since
\begin{eqnarray*}
&&\sup_{\Bbeta_{\lambda|t_{0}} \in D } \left | \sum_{i=1}^{n} G^{-1}(Y_i) [I\{t_{0} - \exp(\Bbeta'_{t_{0}}\BZ_{i}) \leq Y_{i} \leq t_0, \Delta_{i}=1\} \right. \\
 &&\left. - P\{t_{0} - \exp(\Bbeta'_{t_{0}}\BZ_{i}) \leq Y_{i} \leq t_0, \Delta_{i}=1\}] \right| = o(n^{1/2 + \epsilon}),
\end{eqnarray*}
and
$$
\sup_{\Bbeta_{\lambda|t_{0}} \in D } \left | \sum_{i=1}^{n} G^{-1}(Y_i) [I(Y_i \leq t_0,\Delta=1)-P(Y_i \leq t_0,\Delta=1)] \right| = o(n^{1/2 + \epsilon}),
$$
it follows that 
\begin{equation} \sup_{\Bbeta_{\lambda|t_{0}} \in D }  \left |\left |n^{-1/2}\BQ_n(\Bbeta_{\lambda|t_{0}}) - n^{-1/2}\tilde{\BQ}_{n}(\Bbeta_{\lambda|t_{0}}) \right | \right | = o(n^{-1/2 + \epsilon}), \ a.s.
\label{eqn:A1}
\end{equation}
Using the nonexistent mean value theorem (MEMVT) for vector-valued function (Feng {\it et al}., 2013) around $\Bbeta^{0}_{\lambda|t_{0}}$ and letting $\Bbeta^{*}_{t_{0}}$ be some point between $\hat{\Bbeta}_{\lambda|t_{0}}$ and $\Bbeta^{0}_{\lambda|t_{0}}$, we have 
\begin{equation} n^{-1/2}\{\tilde{\BQ}_n(\hat{\Bbeta}_{\lambda|t_{0}}) - \tilde{\BQ}_n(\Bbeta^{0}_{\lambda|t_{0}}) \} \approx (\hat{\Bbeta}_{\lambda|t_{0}} - \Bbeta^{0}_{\lambda|t_{0}})^{'}\BA_{n}(\Bbeta^{*}_{t_{0}}),
\label{eqn:A2}
\end{equation}
where $\BA_{n}(\Bbeta)=n^{-1/2}\partial \tilde{\BQ}_{n}(\Bbeta)/\partial \Bbeta=-(1/n)\sum_{i=1}^n f_i(0)\BZ_{i}\BZ_{i}^{\prime}, f_i(0)$ being the probability density function of $\log(t_0-T_i)-\Bbeta'_{t_{0}}\BZ_{i}$, and hence $\BA_{n}(\Bbeta^{0}_{\lambda|t_{0}})$ is nonpositive definite.
From the definition of $\hat{\Bbeta}_{\lambda|t_{0}}$, we know $n^{-1/2}\BQ_n(\hat{\Bbeta}_{\lambda|t_{0}}) = 0$, and so by (\ref{eqn:A1}) $n^{-1/2}\tilde{\BQ}_{n}(\hat{\Bbeta}_{\lambda|t_{0}})$ will converge to 0, almost surely, as $n \rightarrow \infty$. Also with $n^{-1/2}\tilde{\BQ}_n(\Bbeta^0_{\lambda|t_{0}}) = 0$ from (\ref{eqn:A0}) and $\BA_{n}(\Bbeta^{0}_{\lambda|t_{0}})$ being negative definite, equation (\ref{eqn:A2}) gives $\hat{\Bbeta}_{\lambda|t_{0}} \rightarrow \Bbeta^{0}_{\lambda|t_{0}}$, a.s. as $n \rightarrow \infty$. 

\begin{center}
{\bf Appendix B: Asymptotic Normality of $\BQ_n(\Bbeta^0_{\lambda|t_0})$ and $\hat{\Bbeta}_{\lambda|t_0}$}
\end{center}

Recall that we have the estimating equation 
\[
\BQ_n(\Bbeta_{\lambda|t_0})=n^{-1/2}\sum_{i=1}^n \BZ_i\dfrac{I(Y_i\leq t_0,\Delta_i=1)}{\hat{G}(Y_i)}\left[I\big\{\log(t_0-Y_i)\leq \BZ_i'\Bbeta_{\lambda|t_0}\big\}-\lambda\right].
\]
Also define 
\[
\BQ_n^G(\Bbeta_{\lambda|t_0})=n^{-1/2}\sum_{i=1}^n \BZ_i\dfrac{I(Y_i\leq t_0,\Delta_i=1)}{{G}(Y_i)}\left[I\big\{\log(t_0-Y_i)\leq \BZ_i'\Bbeta_{\lambda|t_0}\big\}-\lambda\right].
\]
We first derive the limiting distribution of $\BQ_n(\Bbeta^0_{\lambda|t_0})$, where
\begin{equation}\label{eqn:splitS}
\BQ_n(\Bbeta^0_{\lambda|t_0})=\BQ_n^G(\Bbeta^0_{\lambda|t_0})+\BQ_n(\Bbeta^0_{\lambda|t_0})-\BQ_n^G(\Bbeta^0_{\lambda|t_0}).
\end{equation}
The first item, $\BQ_n^G(\Bbeta^0_{\lambda|t_0})$, clearly follows a zero-mean Gaussian process,
because of the fact that $\mathcal{F}_1=\{\lambda: \zeta_{1i}(\lambda)\}$
is Donsker, where 
$$\Bzeta_{1i}(\lambda)\equiv\BZ_i{I(Y_i\leq t_0,\Delta_i=1)}G^{-1}(Y_i)\left[I\big\{\log(t_0-Y_i)\leq \BZ_i'\Bbeta^0_{\lambda|t_0}\big\}-\lambda\right].$$ The Donsker's property holds because the class of indicator functions is Donsker, and due to the preservation properties of the Donsker's class (Section 9.4, Kosorok, 2008).

Next, let us define $N_i^G(t)=I(Y_i \leq t, \Delta_i=0)$, $R_i(t)=I(Y_i \geq t)$, $r(t)=P(Y \geq t)$, $N^G(t)=\sum_{i=1}^n N_i^G(t)$, $\Lambda^G(t)=n^{-1} \int_0^t r(s)^{-1}dN^G(s)$, and $M_i^G(t)=N_i^G(t)-\int_0^{\infty}R_i(s)d\Lambda^G(s)$ is the martingale process of the censoring time for the $i^{th}$ subject. From Pepe (1991), we have 
$$\sqrt{n}\{\hat{G}(t)-G(t)\}=n^{-1/2}\sum_{i=1}^n IF_i(t)+o_p(1) \mbox{ for } t\in(0,t_0],$$ 
where $IF_i(t)=G(t)\int_0^tr(s)^{-1}dM_i^G(s)$ denotes the influence function of $\hat{G}(t)$ and $\mathcal{F}_G=\{t\in(0,t_0]: IF_i(t)\}$ has been shown to be Donsker (Peng and Fine, 2009). It immediately follows that $$\sqrt{n}\{\hat{G}^{-1}(t)-G^{-1}(t)\}=-n^{-1/2}\sum_{i=1}^n IF_i(t)/G^2(t)+o_p(1).$$

Therefore, the second part of \eqref{eqn:splitS} can be further written as:
\begin{eqnarray*}
&&\BQ_n(\Bbeta^0_{\lambda|t_0})-\BQ_n^G(\Bbeta^0_{\lambda|t_0})\nonumber\\
&=&n^{-1/2}\sum_{j=1}^n \BZ_j \{I(Y_j\leq t_0,\Delta_j=1)\}\{\{\hat{G}^{-1}(Y_j)-{G}^{-1}(Y_j)\}\left[I\big\{\log(t_0-Y_j)\leq \BZ_j'\Bbeta^0_{\lambda|t_0}\big\}-\lambda \right]\nonumber\\
	&=&-n^{-1/2}\sum_{j=1}^n \BZ_j I(Y_j\leq t_0,\Delta_j=1) G^{-2}(Y_j)\left[I\{\log(t_0-Y_j)\leq \BZ_j'\Bbeta^0_{\lambda|t_0}\}-\lambda \right]n^{-1}\sum_{i=1}^n IF_i(Y_j)+o_p(1)\nonumber\\
	&=& n^{-1/2} \sum_{i=1}^n \Bzeta_{2i}(\lambda)+o_p(1),\label{eqn:zeta2}
	\end{eqnarray*}
	where
	$$
	\Bzeta_{2i}(\lambda)=E_{\mathcal{D}_j}\left[-IF_i(Y_j)\BZ_j{I(Y_j\leq t_0,\Delta_j=1)}G^{-2}(Y_j)\left[I\big\{\log(t_0-Y_j)\leq \BZ_j'\Bbeta^0_{\lambda|t_0}\big\}-\lambda\right]\bigg|\mathcal{D}_i\right].
	$$
	and $\mathcal{D}_i=(Y_i,\Delta_i,\BZ_i)$ denotes all observed data from the $i^{th}$ subject.
	Again applying the preservation rule of the Donsker's class, we see that $\mathcal{F}_2=\{\lambda: \Bzeta_{2i}(\lambda)\}$ still maintains the Donsker's properties. Combining these arguments and defining $\Bzeta_i(\lambda)=\Bzeta_{1i}(\lambda)+\Bzeta_{2i}(\lambda)$, we have that
	$$\BQ_n(\Bbeta^0_{\lambda|t_0})=n^{-1/2}\sum_{i=1}^n \Bzeta_i(\lambda)+o_p(1).
	$$
	The right handside follows a zero-mean Gaussian process, with the variance-covariance matrix of 
	$$
	\BGamma_{\lambda|t_{0}}=E[\Bzeta_i(\lambda)\Bzeta_i^{'}(\lambda^{*})],\label{eqn;B1}
	$$
	which can be consistently estimated using the estimated version of $\Bzeta_i(\lambda)$.
	
	To establish the asymptotic linearity and normality of $\hat{\Bbeta}_{\lambda|t_0}$, let us define $N(t)=I(Y \leq t, \Delta=1)$ and 
	$$\Bmu(\Bbeta_{\lambda|t_0})=E\left[\frac{Z N(t_0)}{G(Y)} [I\{Y \geq t_0-\exp(\Bbeta_{\lambda|t_0}'Z)\}-\lambda] \right].$$
	Following similar steps as in Peng and Huang (2008), we have
	\begin{eqnarray*}
-\BQ_n(\Bbeta^0_{\lambda|t_0})&=&\BQ_n(\hat{\Beta}_{\lambda|t_0})-\BQ_n(\Bbeta^0_{\lambda|t_0})=\sqrt{n}\{\Bmu(\hat{\Beta}_{\lambda|t_0})-\Bmu(\Bbeta^0_{\lambda|t_0})\}+o_p(1)\\
&=&\sqrt{n}\BB(\Bbeta^0_{\lambda|t_0})(\hat{\Beta}_{\lambda|t_0}-\Bbeta^0_{\lambda|t_0})+o_p(1),
	\end{eqnarray*}
where $\Bmu(\hat{\Beta}_{\lambda|t_0})-\Bmu(\Bbeta^0_{\lambda|t_0})$ is the expectation of $n^{-1/2}\left(\BQ_n(\hat{\Beta}_{\lambda|t_0})-\BQ_n(\Bbeta^0_{\lambda|t_0})\right)$ and $\BB(\Bbeta^0_{\lambda|t_0})=\partial \Bmu(\Bbeta_{\lambda|t_0})/\partial \Bbeta_{\lambda|t_0}|_{\Bbeta_{\lambda|t_0}=\Bbeta^0_{\lambda|t_0}}.$
	Therefore, 
$$\sqrt{n}\left(\hat{\Beta}_{\lambda|t_0}-\Bbeta^0_{\lambda|t_0}\right)=-n^{-1/2}\sum_{i=1}^n\BB(\Bbeta^0_{\lambda|t_0})^{-1}\Bzeta_{i}(\lambda)+o_p(1),$$ which weakly converges to Gaussian process with the variance-covariance matrix of $$\BB(\Bbeta^0_{\lambda|t_0})^{-1}E[\Bzeta_i(\lambda)\Bzeta_i^{'}(\lambda^{*})] \BB(\Bbeta^0_{\lambda|t_0})^{-1'}.$$

\begin{center}
{\bf Appendix C: Justification for the Perturbation Scheme}
\end{center}

Following the arguments in Jin {\it et al.} (2001) and 
Peng and Huang (2008), it suffices to show that  
\begin{equation}\label{eqn:perturbj}
\sqrt{n}\left({\Beta}^*_{\lambda|t_0}-\hat\Bbeta_{\lambda|t_0}\right)=-n^{-1/2}\sum_{i=1}^n(\xi_i-1)\BB(\Bbeta^0_{\lambda|t_0})^{-1}\Bzeta_{i}(\lambda)+o_p(1),
\end{equation}
where $\{\xi_i\}_{i=1}^n$ are the i.i.d. positive random variables that
satisfy $E(\xi_i)=\mbox{var}(\xi_i)=1$. This can be justified by following the arguments in Appendix B. Specifically, 
the perturbed version of the Kaplan-Meier estimator, $G^*(t)$, satisfies
the following asymptotic representation,
\begin{equation}\label{eqn:Gstarif}
\sqrt{n}\{{G}^*(t)-G(t)\}=n^{-1/2}\sum_{i=1}^n \xi_i \cdot IF_i(t)+o_p(1) \mbox{ for } t\in(0,t_0].
\end{equation}
Define
\[
\BQ_n^*(\Bbeta_{\lambda|t_0})=n^{-1/2}\sum_{i=1}^n \xi_i\BZ_i\dfrac{I(Y_i\leq t_0,\Delta_i=1)}{{G}^*(Y_i)}\left[I\big\{\log(t_0-Y_i)\leq \BZ_i'\Bbeta_{\lambda|t_0}\big\}-\lambda\right],
\]
and
\[
\BQ_n^{G*}(\Bbeta_{\lambda|t_0})=n^{-1/2}\sum_{i=1}^n \xi_i\BZ_i\dfrac{I(Y_i\leq t_0,\Delta_i=1)}{{G}(Y_i)}\left[I\big\{\log(t_0-Y_i)\leq \BZ_i'\Bbeta_{\lambda|t_0}\big\}-\lambda\right].
\]
Using \eqref{eqn:Gstarif} and following the arguments in Appendix B,
we can show that 
\[\BQ_n^*(\Bbeta^0_{\lambda|t_0})-\BQ_n^{G*}(\Bbeta^0_{\lambda|t_0})=n^{-1/2} \sum_{i=1}^n \xi_i\Bzeta_{2i}(\lambda)+o_p(1),\]
which further implies that $\BQ_n^*(\Bbeta^0_{\lambda|t_0})=n^{-1/2} \sum_{i=1}^n \xi_i\Bzeta_{i}(\lambda)+o_p(1)$.

Since $\Bbeta^*_{\lambda|t_0}$ is the root of the estimating equation 
$\BQ_n^*(\Bbeta_{\lambda|t_0})=0$, we have 
$\BQ_n^*(\Bbeta^*_{\lambda|t_0})-\BQ_n^*(\Bbeta^0_{\lambda|t_0})
=-n^{-1/2} \sum_{i=1}^n \xi_i\Bzeta_{i}(\lambda)+o_p(1)$.
We can then utilize the asymptotic linearity in the vicinity of $\Bbeta^0_{\lambda|t_0}$ as well as Taylor expansion to
$$\sqrt{n}\left({\Beta}^*_{\lambda|t_0}-\Bbeta^0_{\lambda|t_0}\right)=-n^{-1/2}\sum_{i=1}^n\xi_i\BB(\Bbeta^0_{\lambda|t_0})^{-1}\Bzeta_{i}(\lambda)+o_p(1),$$
which, when combined with the results in Appendix B, leads to the expression in \eqref{eqn:perturbj}.


\bibliographystyle{plainnat}

\end{document}